\documentclass[
reprint,
superscriptaddress,
amsmath,
amssymb,
aps,prl]{revtex4-1}

\usepackage{graphicx}% Include figure files
\usepackage{dcolumn}% Align table columns on decimal point
\usepackage{bm}% bold math
\usepackage{hyperref}% add hypertext capabilities
\usepackage{xcolor}
\usepackage[mathlines]{lineno}% Enable numbering of text and display math

\begin{document}

%\preprint{APS/123-QED}

\title{Zone folding induced topological insulators in phononic crystals}

\author{Yuanchen Deng}
\affiliation{Department of Mechanical and Aerospace Engineering, North Carolina State University, Raleigh, North Carolina 27695, USA}
\author{Hao Ge}
\affiliation{National Laboratory of Solid State Microstructures and Department of Materials Science and Engineering, Nanjing University, Nanjing, Jiangsu 210093, China}
\author{Yuan Tian}
\affiliation{National Laboratory of Solid State Microstructures and Department of Materials Science and Engineering, Nanjing University, Nanjing, Jiangsu 210093, China}
\author{Minghui Lu}
\email{luminghui@nju.edu.cn}
\affiliation{National Laboratory of Solid State Microstructures and Department of Materials Science and Engineering, Nanjing University, Nanjing, Jiangsu 210093, China}
\author{Yun Jing}
%\email[Authors to whom correspondence should be addressed. Electronic addresses: ]{yjing2@ncsu.edu}
\email{yjing2@ncsu.edu}
\affiliation{Department of Mechanical and Aerospace Engineering, North Carolina State University, Raleigh, North Carolina 27695, USA}

\date{\today}

\begin{abstract}
This letter investigates a flow-free, pseudospin-based acoustic topological insulator. Zone folding, a strategy originated from photonic crystal, is used to form double Dirac cones in phononic crystal. The lattice symmetry of the phononic crystal is broken by tuning the size of the center ``atom" of the unit cell in order to open the nontrivial topological gap. Robust sound one-way propagation is demonstrated both numerically and experimentally. This study provides a flexible approach for realizing acoustic topological insulators, which are promising for applications such as noise control and waveguide design.

%\begin{description}
%\item[Key Words]
%Phononic crystal; topological insulator; double Dirac cone
%\end{description}
\end{abstract}
%43.25.Qp: Radiation Pressure; 43.25.Gf: Standing acoustic wave; resonance; 42.50.Tx : Optical angular momentum and its quantum aspects; 43.25.Jh： Reflection, refraction, interference, scattering, and diffraction of intense sound waves; 43.20.-f: general linear acoustics

%\pacs{XX}% PACS, the Physics and Astronomy
\maketitle

%\tableofcontents

Recent discoveries in condensed matter physics have opened possibilities for topological physics which are characterized by either the quantum hall effect (QHE)~\cite{Laughlin1983,Klitzing1980} or quantum spin Hall effect (QSHE)~\cite{Bernevig2006,Kane2005}. Translating the concept of topological phases~\cite{Hasan2010} to classical waves such as optic~\cite{He2016,Lumer2013,Liang2013,Wu2015,Khanikaev2013,Rechtsman2013a,Peano2015}, acoustic~\cite{He2016Acoustic,Wei2017,Peng2016,Fleury2016,Khanikaev2015,Chen2016,Ni2015,Yang2015,Lu2016,Lu2016ATI,Zhang2017,He2016ATI,Xiao2015}, and elastic waves~\cite{Susstrunk2016,Mousavi2015,Susstrunk2015,Pal2016}, is currently an active area of research. There are several barriers in realizing topological states in acoustics, such as the absence of polarization in longitudinal waves and the difficulty of breaking the time reversal symmetry in Hermitian systems. For acoustic topological Chern insulators, external fields such as circulating fluids have been used to break the time reversal symmetry~\cite{He2016Acoustic,Wei2017,Peng2016,Fleury2016,Khanikaev2015,Chen2016,Ni2015,Yang2015}. The inevitable dynamic instability and flow-induced noise, however, could pose great challenges for experimental demonstration. On the other hand, the intrinsic spin-1/2 fermionic characteristic is the underpinning component of the QSHE for electrons. Direct analogy in acoustics is non-trivial, though, due to the spin-0 nature of acoustic waves. To address this issue, He~\cite{He2016ATI} $et$ $al.$ and Mei~\cite{mei2016pseudo} $et$ $al.$ independently provided a pseudospin approach for designing flow-free acoustic topological insulators based on accidentally formed double Dirac cones in phononic crystal~\cite{Chen2014,Li2014}. However, the double Dirac cone can only be obtained at a fixed filling ratio, which is typically found by a trial-and-error approach. In addition, the accidental double Dirac cones either rely on a high impedance contrast~\cite{He2016ATI} or composite materials~\cite{mei2016pseudo}. While the former condition is very difficult to satisfy in acoustic media other than air, the latter adds substantial complexity to material fabrication.  In photonic crystals, the ``zone folding" mechanism was proposed to form double Dirac cones~\cite{Wu2015}. By expanding the unit cell of the lattice, the Brillouin zone will fold and the high symmetry points $K$ and $K'$ in the original Brillouin Zone are mapped to $\Gamma$ point of the new Brillouin zone, giving rise to double Dirac cones. The similar concept is introduced to acoustics to construct double Dirac cones as well as acoustic topological insulators for robust one-way propagation~\cite{Zhang2017}. The structure proposed in acoustics, however, requires a refractive index higher than the background medium, which is difficult to realize for airborne sound~\cite{Zhang2017}.

This letter investigates a new structure which combines the advantages of existing approaches for actualizing acoustic topological insulators: (1) the acoustic topological insulator can be constructed using common, homogeneous materials; (2) the double Dirac cones can be conveniently attained regardless of what the filling ratio is thanks to the ``zone folding" strategy. In a traditional triangular lattice phononic crystal featured by a lattice constant $a_0$, the unit cell could be chosen as a hexagon possessing a single "atom" at the center (Figs. 1a and b). The unit cell has a six-fold rotational symmetry $C_6$ as well as a translational symmetry $T_{a_0}$ corresponding to the lattice constant $a_0$. Therefore, the band structure for the unit cell has single Dirac cones at high symmetry points $K$ and $K'$ ($K'$ is not shown in Fig. 2a due to the symmetry). Here we modify the unit cells so that the physical domain is expanded by a factor of $\sqrt{3}$, yielding a lattice constant $a=\sqrt{3}a_0$. While the original "atom" remains at the center, there are 6 other ``1/3 atom"s at each corner of the hexagon, which naturally also alters the Brillouin zone. As the size of the unit cell is expanded by $\sqrt{3}$, the size of the Brillouin zone consequently is reduced by the same factor. Additionally, there is a $\pi/3$ rotation from the original orientation and the change of the Brillouin zone will map the single Dirac cones from $K$ and $K'$ to the new Brillouin zone at $\Gamma$ point, forming double Dirac cones (Fig. 1c and Fig. 2b). In our design for simulations and experiments, the new lattice constant $a$ is 10 mm. The radius of the ``atom" $r$ is 0.35 $a$, i.e., 3.5 mm. Note that the existence of double Dirac cones in the expanded unit cell does not depend upon a fixed filling ratio because of the ``zone folding" mechanism ~\cite{Suppl}(supplemental materials part I). The new unit cell has double Dirac cones at $f_0=$ 17800 Hz as shown by the band structure in Fig. 2(b). Since the $C_6$ symmetry is still preserved in the expanded unit cell, there are also single Dirac cones at $K$ and $K'$ of the new Brillouin zone. The purpose of introducing the double Dirac cones is to gain a four-fold degeneracy, which is the key component to form artificial pseudospin-1/2 states.

\begin{figure}
\includegraphics[width=8.5cm]{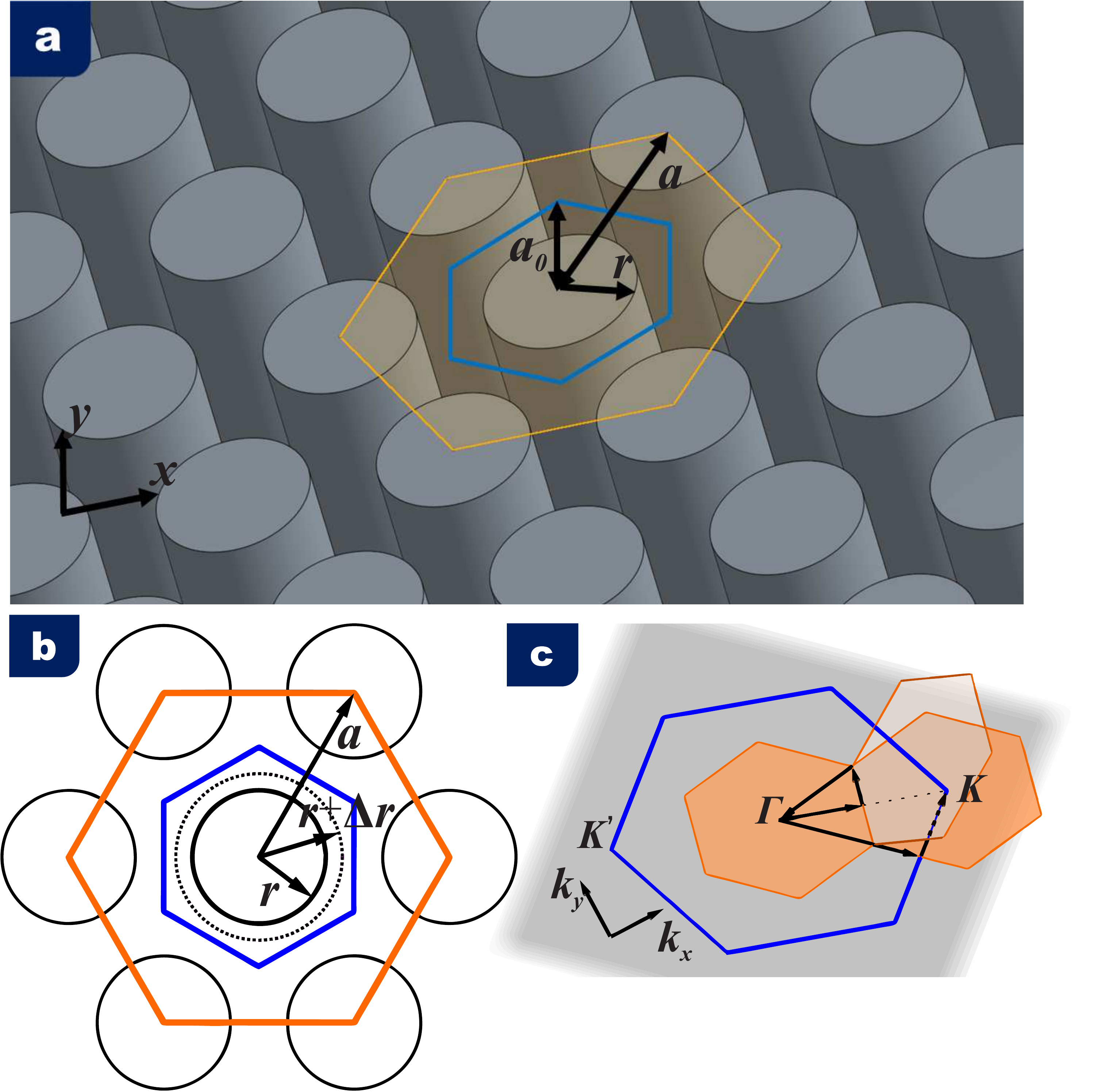}
\caption{\label{fig:1}(a) Structure of the phononic crystal. The rods are periodically arranged with a lattice constant $a_0$. The conventional unit cell of triangular lattice is marked by the blue line whereas the expanded unit cell of triangular lattice is marked by the orange line. (b) Top view of the phononic crystal. $r$ is the radius of the center ``atom" whereas $r+\Delta r$ is the modified radius for breaking the translational symmetry. (c) The conventional Brillouin zone and new Brillouin zone correspond to the two different unit cells. The ``zone folding" mechanism shows how high symmetry points $K$ and $K'$ are mapped to $\Gamma$ point in the new Brillouin zone.}
\end{figure}

Further modification of the unit cell is carried out to break the symmetry in order to open a nontrivial topological band gap at the $\Gamma$ point. In order to do so, the four-fold degeneracy needs to be lifted and separated into two two-fold degenerate modes. To this end, we break the translational symmetry $T_{a_0}$ by either increasing or reducing the radius of the center ``atom" while preserving the $C_6$ symmetry. Figure 1(b) shows the case where the center ``atom" radius is modified by $\Delta r$. In this study, $\Delta r = \pm 1$ mm. Other values can also be used ~\cite{Suppl}(supplemental materials part II). As the radius changes, the $T_{a_0}$ symmetry no longer holds and is replaced by $T_a$ corresponding to the new lattice constant $a$. Figures 2 (c) and (d) show the opening of the bandgap at $\Gamma$ with $\Delta r = 1$ mm and $\Delta r = -1$ mm. In both cases, two two-fold degenerate modes are located on the upper and lower sides of the band gap. Following the convention in quantum mechanics, we can classify these modes to the $p_x/p_y$ and $d_{x^2-y^2}/d_{xy}$ modes according to their eigenmode pressure distributions. In the band structure of the unit cell with $\Delta r = 1$ mm, the $p_x/p_y$ modes are located on the upper side of the bandgap whereas the $d_{x^2-y^2}/d_{xy}$ modes are located on the lower side. This is exactly the opposite of the band structure of the unit cell with $\Delta r = -1$ mm, which is a necessity to bring about band inversion which occurs when $\Delta r = 0$. The degenerate modes that form the two-fold pairs are analyzed by COMSOL Multiphysics 5.2. The material for the rod is steel with a speed of sound of 6010 m/s and a density of 7800 kg/m$^3$. The background medium is air, whose speed of sound is 343 m/s and density is 1.21 kg/m$^3$. It should be pointed out that the material of the rod does not have to be acoustically rigid. This is critical as it provides great flexibility for designing acoustic topological insulators in different background media~\cite{Suppl}(supplemental materials part III). The symmetries of these degenerate modes highly resemble those of $p$/$d$ electrons and are featured with pseudospins of the sound intensity ~\cite{He2016ATI}. As an analogue to the intrinsic spin of electrons in topological insulators, the pseudospin states of acoustic waves in the unit cell are the foundation for building flow-free acoustic topological insulators.

\begin{figure}
\includegraphics[width=8.5cm]{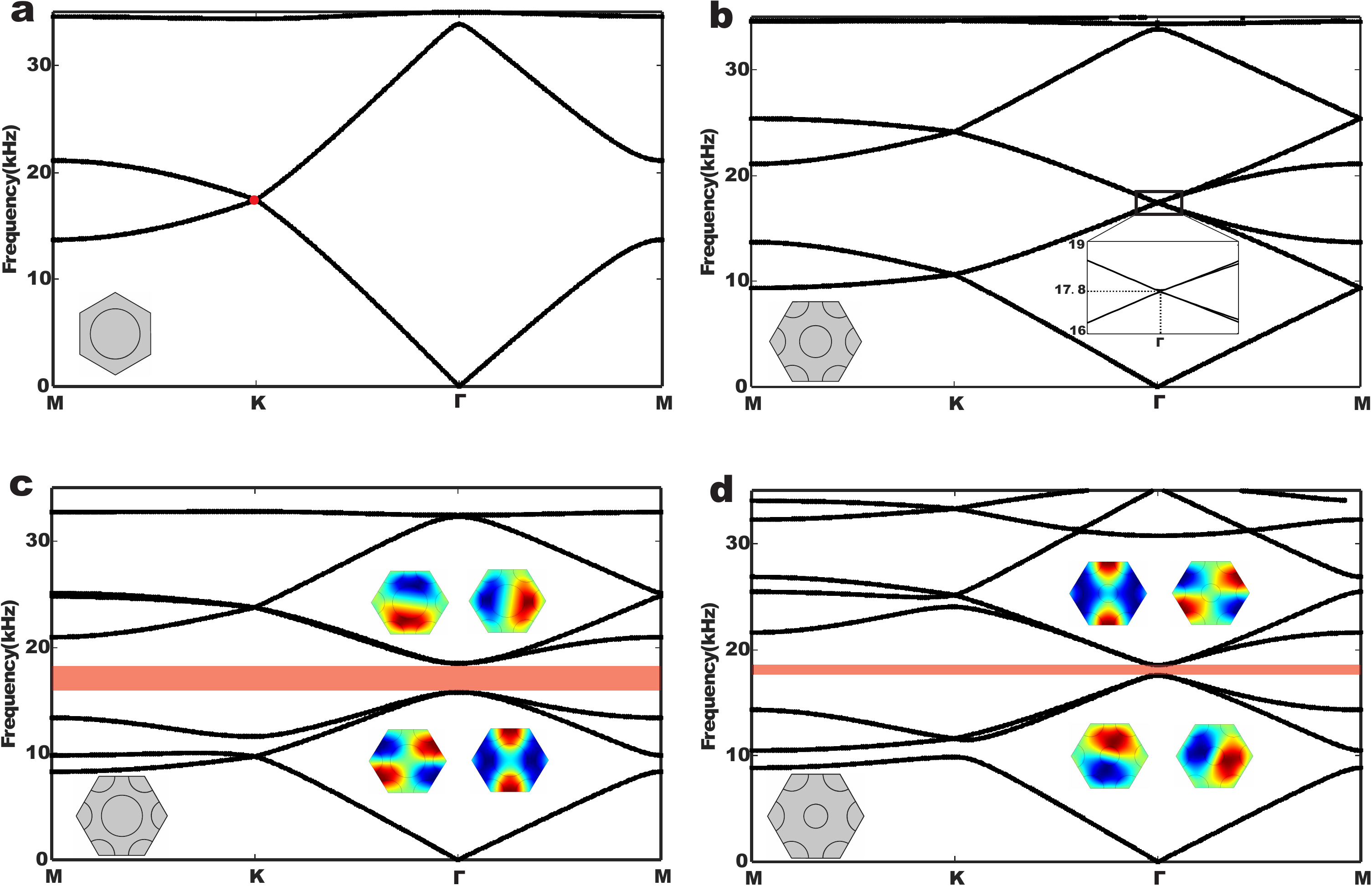}
\caption{\label{fig:2} Band structures for various unit cells. Unit cells are depicted at the bottom left corner in each sub-figure. The rods are assumed to be steel and the background medium is air. (a) Original unit cell with a single ``atom". The single Dirac cone at $K$ is marked by the red dot. (b) Expanded unit cell with double Dirac cones at $\Gamma$. (c)(d) Symmetry broken unit cells with $\Delta r=1$ mm and $\Delta r=-1$ mm, respectively. Topological bandgaps are marked with pink color. $p$/$d$ eigenmodes are shown with their locations separated by the bandgap which indicates the occurrence of band inversion.}
\end{figure}

One important feature of the symmetry-broken unit cell is that the width of the band gap is associated with $\Delta r$. As the radius reduces to a sufficiently small quantity, the lattice becomes a traditional hexagonal lattice which could result in two types of band structures, one with double Dirac cones and one without. For a certain filling ratio, double Dirac cones can be accidentally formed at the $\Gamma$ point of its Brillouin zone~\cite{He2016ATI}. In most cases, however, bandgaps will be preserved.

Topological edge states need to be identified in order to achieve robust one-way propagation. Based upon the symmetry-broken unit cells, we design two types of supercells: one with unit cells that have enlarged center ``atoms" ($\Delta r = 1$ mm, Fig.3a) and one with unit cells that have reduced center ``atoms" ($\Delta r = -1$ mm, Fig.3b).  Bulk bandgaps are found in both cases along the horizontal direction of these supercells, which is resulted from the pseudospin states~\cite{He2016ATI}. To investigate the topologically protected edge state, the two supercells are joined vertically and the band structure is shown in Fig. 3c. The upper supercell has smaller center ``atoms" while the lower one has larger center ``atoms". Non-center ``atoms" are of the same size. The observed topological edge states from FEM simulations between 17450 Hz and 18500 Hz (a minigap ranging from 17713 Hz to 17726 Hz exists though~\cite{He2016ATI}) confirm the existence of the two opposite pseudospin states (spin+ and spin-) on the interior boundary between the two supercells.

\begin{figure}
\includegraphics[width=8.5cm]{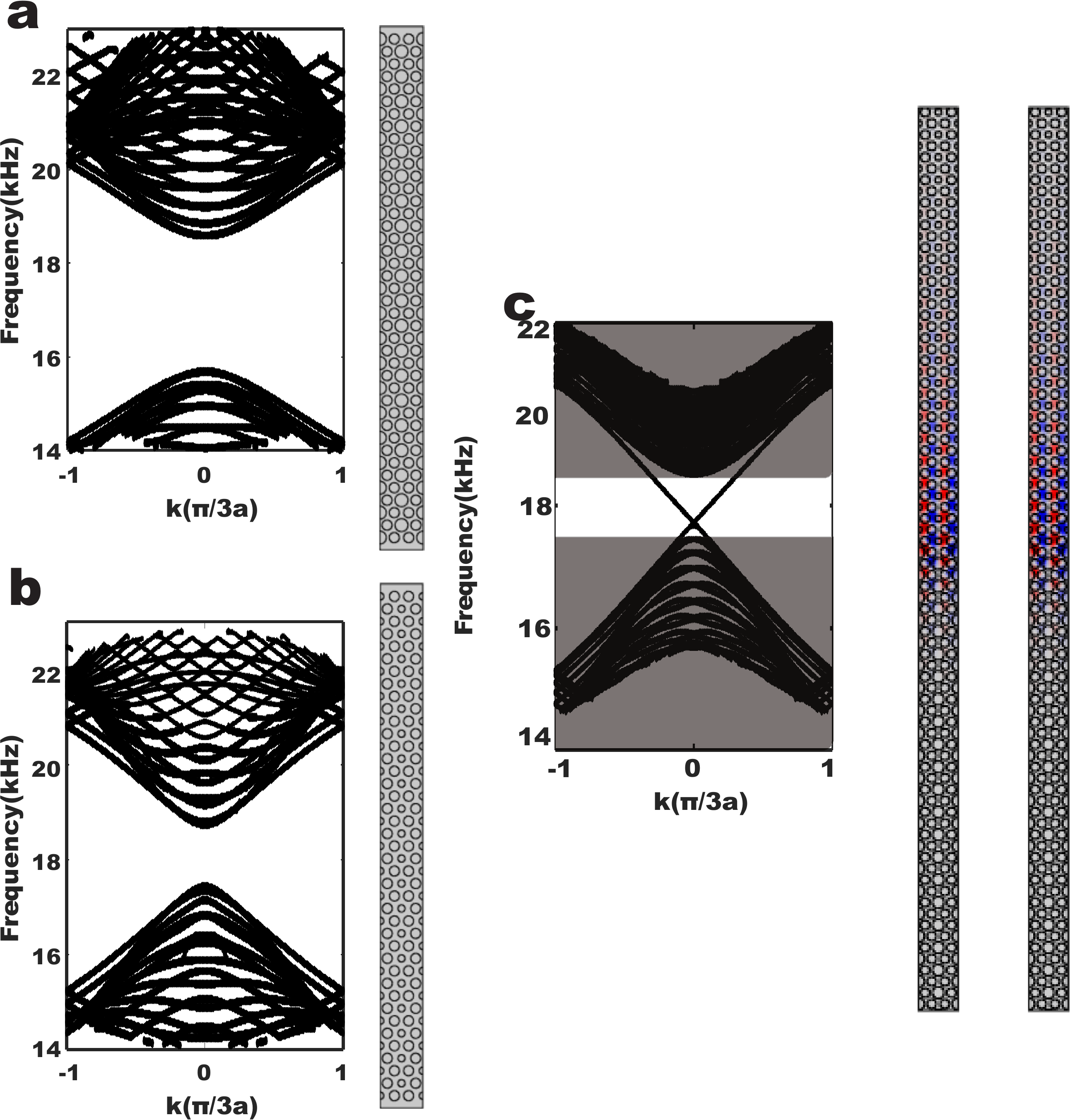}
\caption{\label{fig:3}(a)(b) Band structure of the supercells with enlarged or reduced center "atom"s with the schematic of the supercells shown on the right side. (c) Band structures when the two supercells are joined together. One pair of edge states at $f_0 = 17800$ Hz (arbitrarily chosen within the topological edge state frequency band) are shown on the right.}
\end{figure}

The final step is to both numerically and experimentally observe the hallmark of topological insulators, i.e., symmetry protected one-way propagation. To this end, we construct acoustic topologically protected channels with non-spin-mixing defects sandwiched by phononic crystals consisting of the two types of unit cells. A finite-size sound source is placed at the left port of the channel whereas a microphone is located at the exit of the channel. Finite element method simulations are first performed to model the wave propagation through the phononic crystal. Two cases defined by the channel topology are studied to showcase the one-way propagation phenomenon, i.e., with a cavity and a ``V" shape bending. Results for more complicated defects can be found in the supplemental materials part IV~\cite{Suppl}. The simulated acoustic pressure and intensity distributions at a frequency (18400 Hz was chosen) within the topologically protected edge state frequency band show good transmission even with the presence of defects (Figs. 4a and b). The transmission spectra shown in Fig. 5a demonstrate high sound transmission within a frequency band approximately from 17800 Hz to 18500Hz, slightly narrower than the predicted  frequency band, i.e., 17450 Hz - 18500 Hz (highlighted by the light region). This has happened possibly due to the reflection at the incident boundary (poor impedance match) and the finite size of the topological insulator.

For comparison purposes, an ordinary waveguide in the same phononic crystal is constructed and wave propagation is simulated with the same input (Fig. 4c, d and Fig. 5b). Similarly, we introduce two sets of defects, a cavity and a ``V" shape bending, into the ordinary waveguide. In contrast to the topologically protected channels, the sound transmission spectra here show considerable sound transmission loss in the frequency range of interest, especially for the bending defect case. The distributions of acoustic pressure and intensity in ordinary waveguides indicate that acoustic transmission in ordinary waveguides is considerably more sensitive to defects.

\begin{figure}
\includegraphics[width=8.5cm]{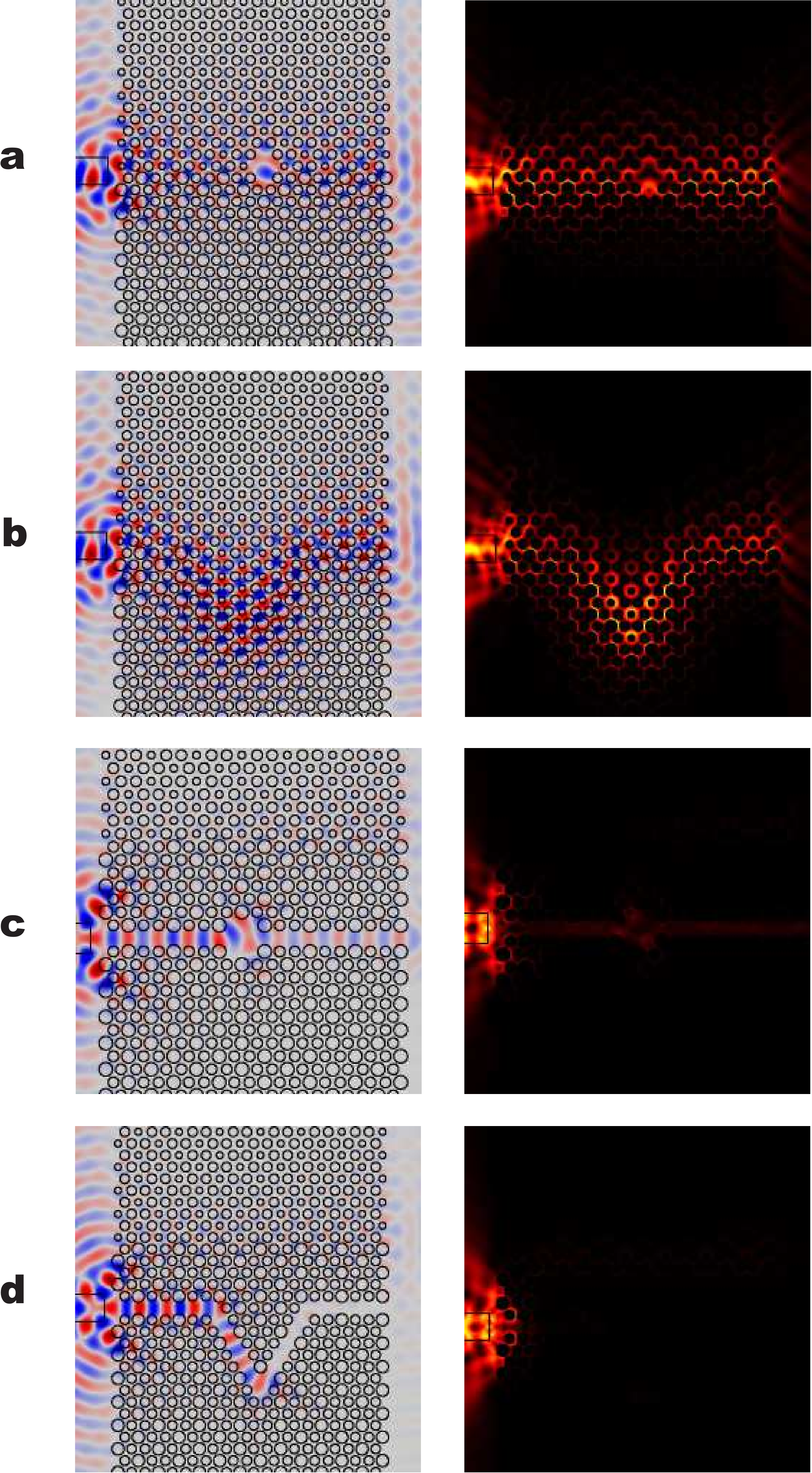}
\caption{\label{fig:4}(a)(b) Acoustic pressure (left) and intensity (right) distributions in the topological insulator with a cavity and a ``V" shape bending.(c)(d) Acoustic pressure and intensity distributions in the ordinary waveguide with the same defects. Considerably lower sound transmissions are observed in the ordinary phononic crystal waveguides. Results are shown at 18400 Hz.}
\end{figure}

\begin{figure}
\includegraphics[width=8.5cm]{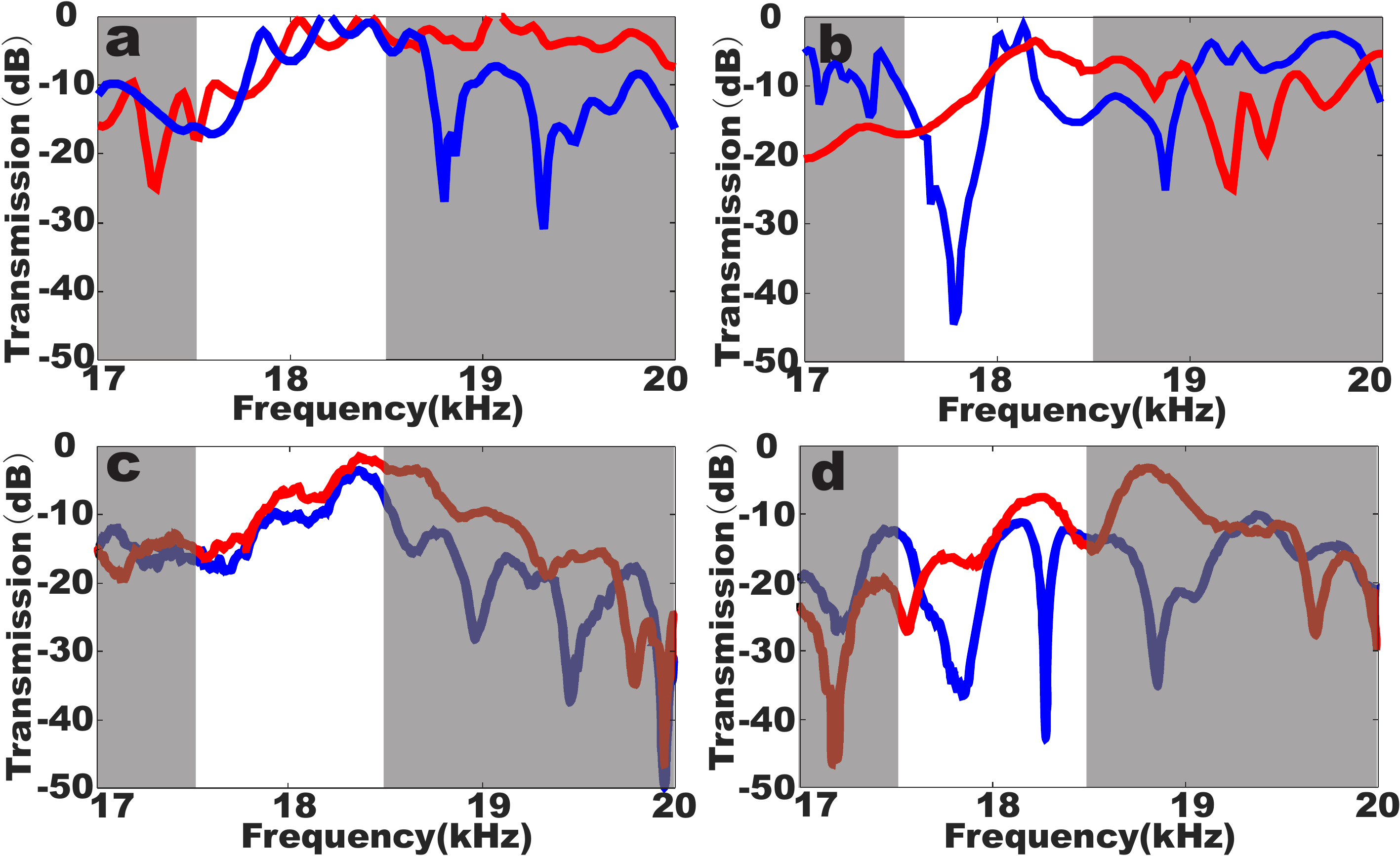}
\caption{\label{fig:5} (a)(b) Simulated transmission spectra with two classes of defects in topological acoustic channels and ordinary waveguide. Red curves represent the cavity case and blue curves represent the bending case. (c)(d) Experimental results of the transmission spectra corresponding to (a) and (b), respectively. The topologically protected edge state frequency range is indicated by the light region.}
\end{figure}

To validate the simulation, experiments are carried out and the measured sound transmission spectra are shown in Figs. 5c and d, in comparison with Figs. 5a and b. Steel rods are used for constructing the phononic crystal, which is consistent with the simulations. More details of the experimental setup can be found in the supplemental materials~\cite{Suppl}(supplemental materials part V). Reasonably good agreement can be observed between the simulation and measurement results, both showing the robustness of one-way propagation in the acoustic topological insulator as well as the ineffective sound transmission in the ordinary waveguides. The discrepancy between the simluation and measurement could stem from the loss in the system, which is not considered in the simluation, as well as fabrication errors and acoustic source imperfection.

In conclusion, the acoustic topological insulator introduced here is designed with the ``zone folding" mechanism which could form double Dirac cones without the constraint on the filling ratio or lattice constant of the phononic crystal. The two-fold degenerate pairs are created from the four-fold degeneracy at the $\Gamma$ point as the radius of the center ``atom" is modified in order to break the translational symmetry. The lattice constructed by two types of symmetry broken unit cells is designed in order to generate topologically protected edge states, which give rise to robust one-way propagation that cannot be observed in phononic crystal-based ordinary waveguides. This unique design addresses several challenges in previous approaches for realizing acoustic topological insulators. While this study focuses on airborne sound, the design strategy can be readily applied to designing underwater acoustic topological insulators operating at ultrasound frequency, which will significantly expand the application range of acoustic topological insulators to areas such as medical ultrasound and sonar devices.

The work was jointly supported by the National Basic Research Program of China (Grant No. 2012CB921503, 2013CB632904 and 2013CB632702) and the National Nature Science Foundation of China (Grant No. 11134006, No. 11474158, and No. 11404164). M.-H.L. also acknowledges the support of the Natural Science Foundation of Jiangsu Province (BK20140019) and the support from the Academic Program Development of Jiangsu Higher Education (PAPD).

Y.D. and H.G. contributed equally to this work.


%\bibliographystyle{apsrev4-1}
%\bibliography{test}

\end{document}